\documentclass{ws-ijmpd}
\usepackage[super,compress]{cite}
\begin{document}

\markboth{Tatekawa}
{Transient from init. based on LPT in N-body simulations III}

%
\catchline{}{}{}{}{}
%

\title{TRANSIENTS FROM INITIAL CONDITIONS BASED ON LAGRANGIAN
PERTURBATION THEORY IN N-BODY SIMULATIONS III: THE CASE OF
GADGET-2 CODE}

\author{TAKAYUKI TATEKAWA}

\address{Department of Social Design Engineering, 
National Institute of Technology, Kochi College,
200-1 Monobe-Otsu, Nankoku, Kochi,
 783-8508, JAPAN\\
esearch Institute for Science and Engineering,
Waseda University, 3-4-1 Okubo, Shinjuku,
Tokyo 169-8555, JAPAN\\
tatekawa@akane.waseda.jp}

\maketitle

\begin{history}
\received{Day Month Year}
\revised{Day Month Year}
\end{history}

\begin{abstract}
In modern cosmology, the precision of the theoretical prediction is increasingly
required. In cosmological $N$-body simulations, the effect of higher-order
Lagrangian perturbation on the initial conditions appears in terms of statistical
quantities of matter density field. We have considered the effect
of third-order Lagrangian perturbation (3LPT) on the initial
conditions, which can be applied to Gadget-2 code. Then, as
statistical quantities, non-Gaussianity
of matter density field has been compared between cases of different
order perturbations for the initial conditions. Then, we demonstrate
the validity of the initial conditions with 
second-order Lagrangian perturbation (2LPT).
\end{abstract}


\ccode{02.60.Cb, 02.70.-c, 04.25.-g, 98.65.Dx}



\section{Introduction}
 \label{sec:Intro}

Based on recent observations, refinement of the cosmological scenario
is under progress~\cite{2dF,SDSS,LSST,DES,Euclid}. 
For example, galaxy surveys not only
present large-scale structures but also evolution of
such structures in the Universe~\cite{LSST,DES,Euclid}. 
As the evolution of large-scale structures is clarified, 
various dark energy models~\cite{Peebles2003,Copeland2006},
 which explain
the acceleration of the cosmic expansion, would be restricted.

As one of the useful methods to restrict cosmological models
such as dark energy scenario, cosmological $N$-body simulations
have been applied~\cite{Miyoshi, Klypin83, Efstathiou:1985re, P3M, Bertschnger1998, Gadget}, which
describe
the evolution of nonlinear structures such as cluster of galaxies.
Because cosmological $N$-body simulations include the cosmic expansion,
evolution of the nonlinear structures would be affected by the dark energy models.
By the comparison between observations and the predictions by the
cosmological $N$-body simulations, we can verify the validity
of dark energy models. 

For precise verification of cosmological models such as dark energy model,
precise simulations are required. We focus on the initial condition
for cosmological $N$-body simulations, where Lagrangian linear perturbation,
i.e., Zel'dovich approximation has been used for a long time.
However, although Zel'dovich approximation describes the evolution
of quasi-nonlinear density field, because it is described
by linear perturbation, initial conditions fail to take into account higher-order growing modes
\cite{Scoccimarro:1997gr,Valageas:2002}.
Recently, the effect of second-order Lagrangian perturbation (2LPT) on
the initial condition for cosmological simulation has been studied~\cite{Crocce2006}, which is
manifested 
in the nonlinear structure at low-z era.

We investigated the effect of third-order Lagrangian perturbation (3LPT)
on the initial condition for cosmological simulation~\cite{Tatekawa07, Tatekawa14}.
In the previous studies, we used $P^3 M$ code for cosmological simulations
~\cite{P3M}.
Although the execution speed of the simulation code is fast, $P^3 M$ code
can be applied for structure formation of cold dark matter only.

Gadget-2~\cite{Gadget} is
a well-known code for cosmological $N$-body/SPH simulation and can consider not only cold dark matter
but also baryonic matter. 
The code can be executed on parallel computers with distributed
memory. Therefore, huge simulations can be implemented with this code
~\cite{Springel2005}. Hence, we have developed 3LPT initial condition code applicable to Gadget-2, which
would be quite useful for various analyses considering several situations.

The effect of 3LPT on the initial condition is analysed in terms of 
statistical quantities for matter density field. Even if the initial
condition is given by Gaussian distribution, the matter density field
shows non-Gaussian distribution during nonlinear evolution.
If the initial condition is set at $z=49$, the difference of
the non-Gaussianity between the cases of 2LPT and 3LPT
appears about $0.5 \%$. When we choose initial time at
$z=99$, that difference almost disappear.
From these results, we evaluate the effect of 3LPT
for initial conditions and clarify the validity of 
initial conditions with 2LPT.

This paper is organized as follows.
In Sec.~\ref{sec:NonL}, we present Lagrangian perturbations
valid up to the third-order. Then, we discuss
the methods and results of the numerical simulations 
in Sec.~\ref{sec:NumRes}. In this section, we also introduce
statistical quantities for matter distribution. Finally,
Sec.~\ref{sec:Summary} presents the conclusions.


\section{Lagrangian perturbations} 
\label{sec:NonL}

\subsection{basic equations}

In this section, we briefly introduce Lagrangian perturbation.
When the scale of an object is smaller than that of the cosmological
horizon, the description of motion of matter by Newtonian dynamics
is valid. The cosmological expansion is affected by the scale factor
$a$ in basic equations (continuous equation,
Euler's equation, and Poisson's equation). The solution $a$ is derived by Friedmann's
equations or alternative equations. We consider
dust fluid, which can ignore the pressure of matter.
In the comoving coordinates, the basic equations 
are described as follows
~\cite{Peebles, Liddle, Coles, Weinberg, Sahni-Coles}:
\begin{eqnarray}
\frac{\partial \delta}{\partial t} + \frac{1}{a} \nabla_x \cdot
\left \{ \bm{v} (1+\delta) \right \} &=& 0 \,,
 \label{eqn:conti} \\
\frac{\partial \bm{v}}{\partial t} + \frac{1}{a}
 \left (\bm{v} \cdot \nabla_x \right ) \bm{v} 
 + \frac{\dot{a}}{a} \bm{v} &=& \frac{1}{a} \tilde{\bm{g}} \,,
 \label{eqn:Euler} \\
\nabla_x \cdot \tilde{\bm{g}} &=& - 4 \pi G \bar{\rho} a \delta \,,
\end{eqnarray}
where $\bar{\rho}$ represents background matter density. 
The density fluctuation $\delta$ is defined as
\begin{equation}
\delta \equiv \frac{\rho - \bar{\rho}}{\bar{\rho}} \,.
\end{equation}
$\bm{v}$ denotes peculiar velocity.

In Eulerian perturbation theory, the density fluctuation $\delta$ is
regarded as a perturbation. On the other hand, in Lagrangian perturbation
theory, displacement from a homogeneous distribution is considered
as a perturbation~\cite{Sahni-Coles,ZA,Bernardeau02,Tatekawa04R}.
\begin{equation} \label{eqn:x=q+s}
\bm{x} = \bm{q} + \bm{s} (\bm{q}, t) \,,
\end{equation}
where $\bm{x}$ and $\bm{q}$ represent comoving Eulerian coordinates
and Lagrangian coordinates, respectively. $\bm{s}$ denotes the
displacement vector, which is regarded as a perturbation 
quantity. By the Lagrangian perturbation (\ref{eqn:x=q+s}),
we can solve continuous equation (\ref{eqn:conti}) exactly.
\begin{equation} \label{eqn:delta-Jacobian}
\delta = 1 - J^{-1} \,, J \equiv \det
 \left ( \frac{\partial x_i}{\partial q_j} \right ) \,,
\end{equation}
$J$ refers to the Jacobian of the coordinate transformation from 
Eulerian $\bm{x}$ to Lagrangian $\bm{q}$. Therefore, when we
derive the solution of Lagrangian displacement $\bm{s}$, we
can determine the evolution of the density fluctuation.

The peculiar velocity is given as
\begin{equation}
\bm{v} = a \dot{\bm{s}} \,.
\end{equation}
We introduce Lagrangian time derivative
\begin{equation} \label{eqn:Lagrangian-t}
\frac{{\rm d}}{{\rm d} t} \equiv \frac{\partial}{\partial t}
 + \frac{1}{a} \bm{v} \cdot \nabla_x \,.
\end{equation}
Taking the divergence and rotation of Euler's equation
(\ref{eqn:Euler}), we obtain evolution equations for the 
Lagrangian displacement.
\begin{eqnarray}
\nabla_x \cdot \left (\ddot{\bm{s}} + 2 \frac{\dot{a}}{a}
 \dot{\bm{s}} \right ) &=& - 4 \pi G \bar{\rho} (J^{-1}-1) \,, 
 \label{eqn:s-div} \\
\nabla_x \times \left (\ddot{\bm{s}} + 2 \frac{\dot{a}}{a}
 \dot{\bm{s}} \right ) &=& \bm{0} \label{eqn:s-rot} \,.
\end{eqnarray}
Here, superscript dot $\dot{\bm{s}}$ refers to Lagrangian time derivative
(\ref{eqn:Lagrangian-t}).
\begin{equation}
\dot{\bm{s}} = \frac{{\rm d} \bm{s}}{{\rm d} t} \,.
\end{equation}

To solve the Lagrangian perturbative equations, we decompose
the Lagrangian perturbation into its longitudinal and transverse
mode.
\begin{equation}
s_i = \psi_{,i} + \zeta_i \,,
\end{equation}
\begin{equation}
\zeta_{i,i} = 0 \,.
\end{equation}
where subscript ${}_{,i}$ denotes the Lagrangian spatial derivative.

We convert the spatial derivative from Eulerian coordinates
to Lagrangian coordinates in equations (\ref{eqn:s-div}) and
(\ref{eqn:s-rot}).
\begin{eqnarray}
\frac{\partial}{\partial x_i} &=& \frac{\partial}{\partial q_i}
 - s_{j, i} \frac{\partial}{\partial x_j} \nonumber \\
&=& \frac{\partial}{\partial q_i}
 - s_{j, i} \frac{\partial}{\partial q_j} 
 + s_{j, i} s_{k, j} \frac{\partial}{\partial x_k} \nonumber \\
&=& \frac{\partial}{\partial q_i}
 - s_{j, i} \frac{\partial}{\partial q_j} 
 + s_{j, i} s_{k, j} \frac{\partial}{\partial q_k}
 + \cdots \,. 
\end{eqnarray}
where comma indicates Lagrangian spatial derivative.
\[
s_{j, i} = \frac{\partial s_j}{\partial q_i} \,.
\]
%


\subsection{Lagrangian perturbative equations} \label{sec:Lag}
In this subsection, we derive Lagrangian perturbative equations.
The Lagrangian perturbation can be divided into temporal and
spatial parts.
\begin{eqnarray}
\psi &=& g^{(1)} \psi^{(1)} + g^{(2)} \psi^{(2)} + g^{(3)} \psi^{(3)} + \cdots \,, \\
\zeta_i &=& g^{(1T)} \zeta_i^{(1)} + g^{(2T)} \zeta_i^{(2)}
 + g^{(3T)} \zeta^{(3)} + \cdots \,,
\end{eqnarray}
where superscript ${}^{(n)}$ denotes $n$-th order perturbation.

For the first-order perturbation, i.e., Zel'dovich approximation
~\cite{ZA},
the differential equation for the temporal part is given as follows:
\begin{equation}
\ddot{g}^{(1)} + 2 \frac{\dot{a}}{a} \dot{g}^{(1)}
 - 4 \pi G \bar{\rho} g^{(1)} =0 \,.
\end{equation}
When we consider only the growing mode of the temporal parts
and set the temporal parts at the initial condition by
$g^{(1)} (t_{\rm ini})=1$, the Lagrangian displacement is described
by the density fluctuation.
\begin{equation}
\psi_{,ii}^{(1)} (\bm{q}) = - \delta(\bm{q}) \,.
\end{equation}
In other words, the first-order perturbation would be derived
by the initial density fluctuation.

If the primordial vorticity does not exist, the vorticity
never appears during evolution. Even if the primordial vorticity exists,
the transverse mode in the first-order perturbation does not
have a growing solution. Therefore, hereafter we ignore
the transverse mode in the first-order perturbation.

The second-order perturbation is also divided into spatial and
temporal parts~\cite{Bouchet92,Buchert93,Munshi:1994zb}.
The equations are described as follows:
\begin{eqnarray}
\psi_{,ii}^{(2)} &=& \frac{1}{2} \left \{
 \psi_{,ii}^{(1)} \psi_{,jj}^{(1)} - \psi_{,ij}^{(1)}
 \psi_{,ij}^{(1)} \right \} \,,  \label{eqn:S-2}  \\
\ddot{g}^{(2)} + 2 \frac{\dot{a}}{a} \dot{g}^{(2)}
 - 4 \pi G \bar{\rho} g^{(2)} &=& - 4 \pi G \bar{\rho} \left \{ g^{(1)} \right \}^2 \,.
\end{eqnarray}

The third-order perturbation is derived from triplet term
of the first-order perturbation and cross-section of
the first- and the second-order perturbation
~\cite{Buchert94,Bouchet95,Catelan95}.
\begin{eqnarray}
\psi_{,ii}^{(3a)} &=& \det \left ( \psi_{,ij}^{(1)} \right ) \nonumber \\
 &=& \frac{1}{6} \psi_{,ii}^{(1)} \psi_{,jj}^{(1)} \psi_{,kk}^{(1)}
 - \frac{1}{2} \psi_{,ii}^{(1)} \psi_{,jk}^{(1)} \psi_{,jk}^{(1)}
 + \frac{1}{3} \psi_{,ij}^{(1)} \psi_{,jk}^{(1)} \psi_{,ki}^{(1)} \,, \label{eqn:S-3a} \\
\psi_{,ii}^{(3b)} &=& \frac{1}{2} \left \{
 \psi_{,ii}^{(1)} \psi_{,jj}^{(2)} - \psi_{,ij}^{(1)}
 \psi_{,ij}^{(2)} \right \} \label{eqn:S-3b} \,,
\end{eqnarray}
\begin{eqnarray}
\ddot{g}^{(3a)} + 2 \frac{\dot{a}}{a} \dot{g}^{(3a)}
 - 4 \pi G \bar{\rho} g^{(3a)} &=& - 8 \pi G \bar{\rho} 
 \left ( g^{(1)} \right )^3 \,, \\
\ddot{g}^{(3b)} + 2 \frac{\dot{a}}{a} \dot{g}^{(3b)}
 - 4 \pi G \bar{\rho} g^{(3b)} &=& - 8 \pi G \bar{\rho} g^{(1)}
 \left \{ g^{(2)} - \left (g^{(1)} \right )^2 \right \} \,,
\end{eqnarray}

Even if we do not consider transverse mode in the first-order
perturbation, the transverse mode in the third-order appears
~\cite{Sasaki1998}.
\begin{eqnarray}
\ddot{g}^{(3T)} + 2 \frac{\dot{a}}{a}
\dot{g}^{(3T)} &=& 4 \pi G \rho_b \left ( g^{(1)} \right )^3 \,, \\
- \nabla^2 \zeta_i^{(3)} &=& \left ( \psi_{,il}^{(1)} \psi_{,kl}^{(2)}
 - \psi_{,kl}^{(1)} \psi_{,il}^{(2)} \right )_{,k} \,.
\end{eqnarray}
Because of Kelvin's circulation theorem, the transverse mode
in the third-order perturbation does not imply vorticity.
In this paper, the effect of the transverse mode in the third-order
perturbation is also analyzed.

In $\Lambda$CDM model, the early stage in structure formation
is a matter dominant era. Because the effect of cosmological constant
seems negligible, the cosmic expansion would be approximated by
the solution of Einstein-de Sitter Universe model.
\begin{equation}
a(t) \propto t^{2/3} \,.
\end{equation}
Under this assumption, the perturbative solutions become as follows:
\begin{eqnarray}
g^{(1)}(t) &=& t^{2/3} \,, \\
g^{(2)}(t) &=& -\frac{3}{7} t^{4/3} \,, \\
g^{(3a)} (t) &=& \frac{10}{21} t^2 \,, \\
g^{(3b)} (t) &=& -\frac{1}{3} t^2 \,,
\end{eqnarray}

Bouchet {\it et al.} \cite{Bouchet95} derived approximation formula of temporal parts
for $\Lambda$CDM model. They introduced the logarithmic derivative
of the growth factors
\begin{equation}
f_n = \frac{a}{g^{(n)}} \frac{{\rm d} g^{(n)}}{{\rm d} a} \,.
\end{equation}
When the Universe is in the matter dominant era ($\Omega_m \simeq 1$), the
formula becomes
\begin{eqnarray}
f_1 \simeq \Omega_m^{6/11} \, &,& f_2 \simeq 2 \Omega_m^{153/286} \,, \\
f_{3a} \simeq 3 \Omega_m^{146/275} \, &,& f_{3b} \simeq 3 \Omega_m^{9481/17875} \,.
\end{eqnarray}

For the case of $0.1 \le \Omega_m \le 1$, the formula becomes
\begin{eqnarray}
f_1 \simeq \Omega_m^{5/9} \, &,& f_2 \simeq 2 \Omega_m^{6/11} \,, \\
f_{3a} \simeq 3 \Omega_m^{13/24} \, &,& f_{3b} \simeq 3 \Omega_m^{13/24} \,.
\end{eqnarray}
2LPT\_IC code~\cite{Crocce2006} was implemented with the above formula.


\section{Cosmological Simulations} \label{sec:NumRes}
\subsection{Setup of initial conditions}
For precise cosmological simulations, we set up precise initial conditions.
For execution of Gadget-2, we developed a convert code of the initial
conditions from ZA to 3LPT. The convert code is described 
in Sec~\ref{sec:Lag}. 
In 3LPT, analysis is performed separately for the presence or absence of the transverse
mode. Hereafter, those without the transverse mode are described as ``3LPT L'',
and those with the transverse mode are described as ``3LPT L+T''.

In this study, we set the initial condition at the redshift $z_{\rm ini}=49$.
Because the effect of the cosmological constant is negligible,
the temporal components in the convert code are given by solutions
of the Einstein--de Sitter model.

We set the cosmological parameters as shown in Table~\ref{tab:cosmo-param}.
The parameters of simulations are shown in Table~\ref{tab:Gadget-param}.
These parameters are sample values in the Gadget-2 code, which
are slightly different from the recent observation~\cite{Planck2018}.

\begin{table}[ph]
\tbl{Cosmological parameters
in the simulations. }
{\begin{tabular}{lc} \toprule
$\Omega_M$ & $0.25$ \\
$\Omega_{\Lambda}$ & $0.75$ \\
$\Omega_b$ & $0.04$  \\
$H_0$ [km/s/Mpc] & $70$ \\
$\sigma_8$ & $0.8$ \\
$n_0$ & $1.0$ \\ \botrule
\end{tabular} \label{tab:cosmo-param}}
\end{table}

\begin{table}[h]
\tbl{Parameters
in the cosmological simulation code. }
{\begin{tabular}{lc} \toprule
Initial time $z_{in}$ & $49$ \\
Box size $L$ & $100 h^{-1}$ [Mpc] \\
Number of particles $N$ & $256^3$ \\
$h$ & $0.7$ \\
Softening Length & $0.25 h^{-1}$ [Mpc] \\ \botrule
\end{tabular} \label{tab:Gadget-param}}
\end{table}

Gadget-2 can be
executed on many cores by OpenMPI. The simulation code was executed on Linux PC
(CentOS 7.7, Core i9 7960X, RAM 64GB), 
using which the simulation can be executed for approximately 5 hours
per one sample.

Our code was converted from the initial condition generated by ZA to that with
2LPT and 3LPT.
Because our code cannot be executed
parallelly, we cannot apply the code for huge simulations.
For the case of $N=256^3$, the code requires about $2$ GB memory.
Especially in the calculation of the transverse mode in the
third-order perturbation, the code occupies a large amount of memory.

In this simulation, we generated 10 initial conditions for each case.
For the analysis of time evolution, we selected 11 time slices ($z=10, 9, \cdots, 1, 0$).
and compared the density distributions. The density field was smoothed over
the scale $R$ using the cloud-in-a-cell (CIC) algorithm.
The smoothing scale was set as $1 ~h^{-1}, 2 ~h^{-1} \mbox{[Mpc]}$.


\subsection{Non-Gaussianity}
For a detailed analysis, we apply the non-Gaussianity of the density fluctuation.
Even if the primordial density fluctuation is generated by Gaussian distribution,
The non-Gaussianity of the density fluctuation would appear through nonlinear
evolution. For the analysis of the non-Gaussianity, we introduced higher-order
statistical quantities:
\begin{eqnarray*}
\mbox{skewness} &:& \gamma = \frac{ \left < \delta ^3 \right > }
{\sigma^4} \,, \\
\mbox{kurtosis} &:& \eta = 
\frac{\left < \delta^4 \right > - 3 \sigma^4}{\sigma^6} \,,
\end{eqnarray*}
where $\sigma$ means dispersion of the density fluctuation.
\begin{equation}
\sigma = \sqrt{\left <\delta^2 \right >} \,.
\end{equation}
In the weakly nonlinear stage, these statistical quantities were
derived by second-order perturbation theory~\cite{Peebles,Bernardeau02}. 

\subsection{Effect of higher-order perturbations}
We analyze the effect of cosmological simulation 
when 3LPT is included in the initial conditions.
In 3LPT, the effect of the transverse mode is extremely small, 
so the effect of the transverse mode is not discussed in this subsection.
The effect of the transverse mode is analyzed in Section~\ref{sec:3LPT-T}.

\begin{figure}[tb]
\centerline{
\includegraphics[height=11cm]{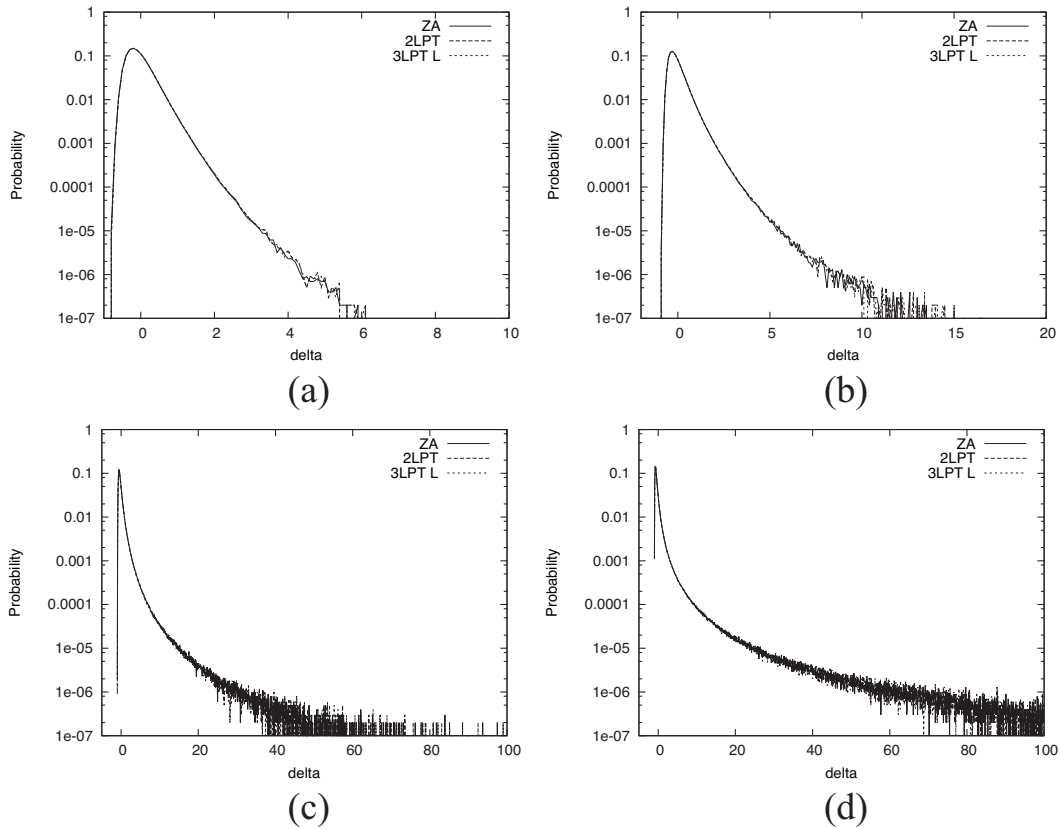}
}
\caption{Distribution function of the density fluctuation
from $N$-body simulation ($R \simeq 1 h^{-1}$ Mpc) with different
initial conditions. 
(a) $z=5$, (b) $z=3$, (c) $z=1$, (d) $z=0$.
The distribution function approaches the log-normal form during the evolution.
}
\label{fig:delta-dist-1M}
\end{figure}

First, we set the smoothing scale as about $1 h^{-1} \mbox{[Mpc]}$.
The distribution function of the density fluctuation is shown in Fig.~\ref{fig:delta-dist-1M}.
It is well-known that the distribution function of the density fluctuation approaches
to log-normal form during the evolution
~\cite{Hamilton1985, Kofman1994, Kayo2001, Ostriker2003}. 
At $z=5$, the effect of higher-order perturbation appeared at the high-density region
$\delta > 10$. During the evolution, the high-density region grows rapidly.
At $z=0$, the distribution functions resemble each other.

In our previous study, we showed the difference of the non-Gaussianity of the density
fluctuation between the initial
conditions given by ZA, 2LPT, and 3LPT. The difference between the cases of
2LPT and 3LPT is about several percent.

\begin{figure}[tb]
\centerline{
\includegraphics[height=5cm]{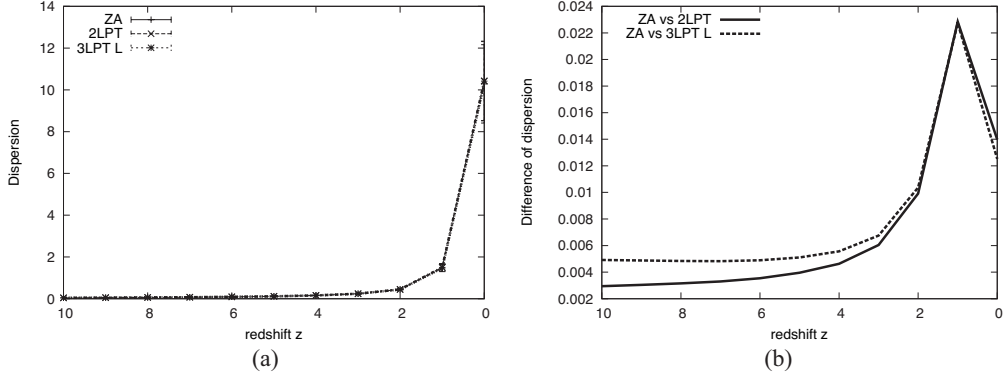}
}
\caption{Dispersion of the density fluctuation
from $N$-body simulation ($R \simeq 1 h^{-1}$ Mpc) with different
initial conditions. (a) Comparison of the dispersion
between the initial conditions. (b) The relative difference of the dispersion
between ZA and other cases.}
\label{fig:dispersion1M}
\end{figure}

First, we show the evolution of the density dispersion with error bars.
The evolution of the density dispersion is shown in Fig.~\ref{fig:dispersion1M}.
At $z=1$, the difference of the density dispersion between the case of ZA
and 3LPT becomes about $2.5 \%$. Furthermore, the difference of
the density dispersion between the case of 2LPT and 3LPT 
is about $0.2\%$. We will notice
the effect of transverse mode in 3LPT later.

\begin{figure}[tb]
\centerline{
\includegraphics[height=5cm]{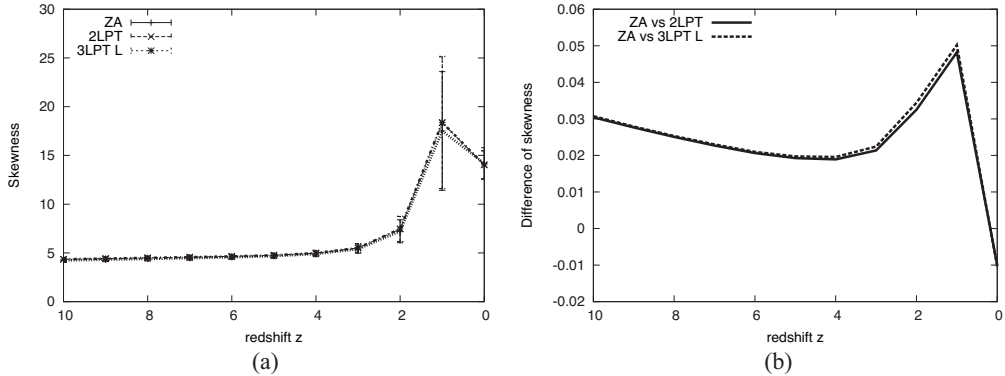}
}
\caption{Skewness of the density fluctuation
from the $N$-body simulation ($R \simeq 1 h^{-1}$ Mpc) with different
initial conditions. (a) Comparison of the skewness
between the initial conditions. (b) The relative difference in the skewness
between ZA and other cases.
}
\label{fig:skewness1M}
\end{figure}

\begin{figure}[tb]
\centerline{
\includegraphics[height=5cm]{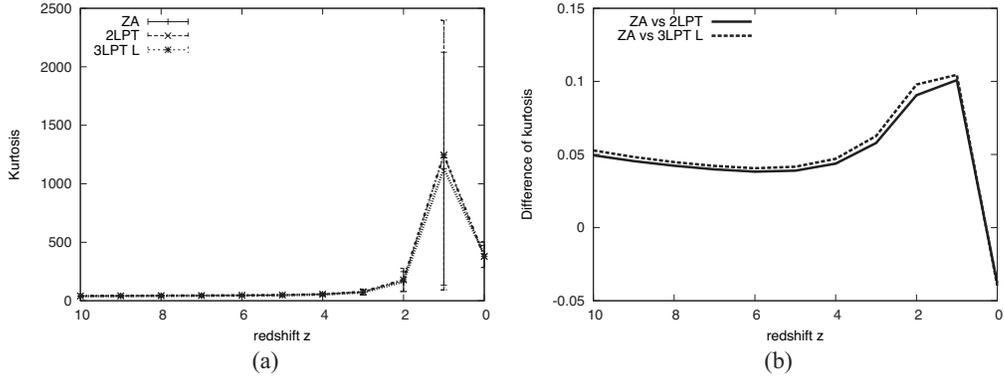}
}
\caption{Kurtosis of the density fluctuation
from the $N$-body simulation ($R \simeq 1 h^{-1}$ [Mpc]) with different
initial conditions. (a) Comparison of the kurtosis
between the initial conditions. (b) The relative difference in the kurtosis
between ZA and other cases.
}
\label{fig:kurtosis1M}
\end{figure}

Then, we show the evolution of the non-Gaussianity in Figs.~\ref{fig:skewness1M}
and \ref{fig:kurtosis1M}.
Because variations among samples are very large, 
in the subsequent analysis, the error bars were omitted.
By comparison between the case of ZA and higher-order perturbations,
the difference in the non-Gaussianity is about $5\%$.
When $z = 0$, the difference in the non-Gaussianity
between the cases of 2LPT and 3LPT becomes very small.
The difference of skewness and kurtosis between the case of
2LPT and 3LPT is about $0.2 \%$ and $0.5 \%$, respectively.
In this analysis, higher-order perturbations in the initial 
conditions affects the difference in the non-Gaussianity 
between models in high-z era ($z \simeq 10$).

\begin{figure}[tb]
\centerline{
\includegraphics[height=10cm]{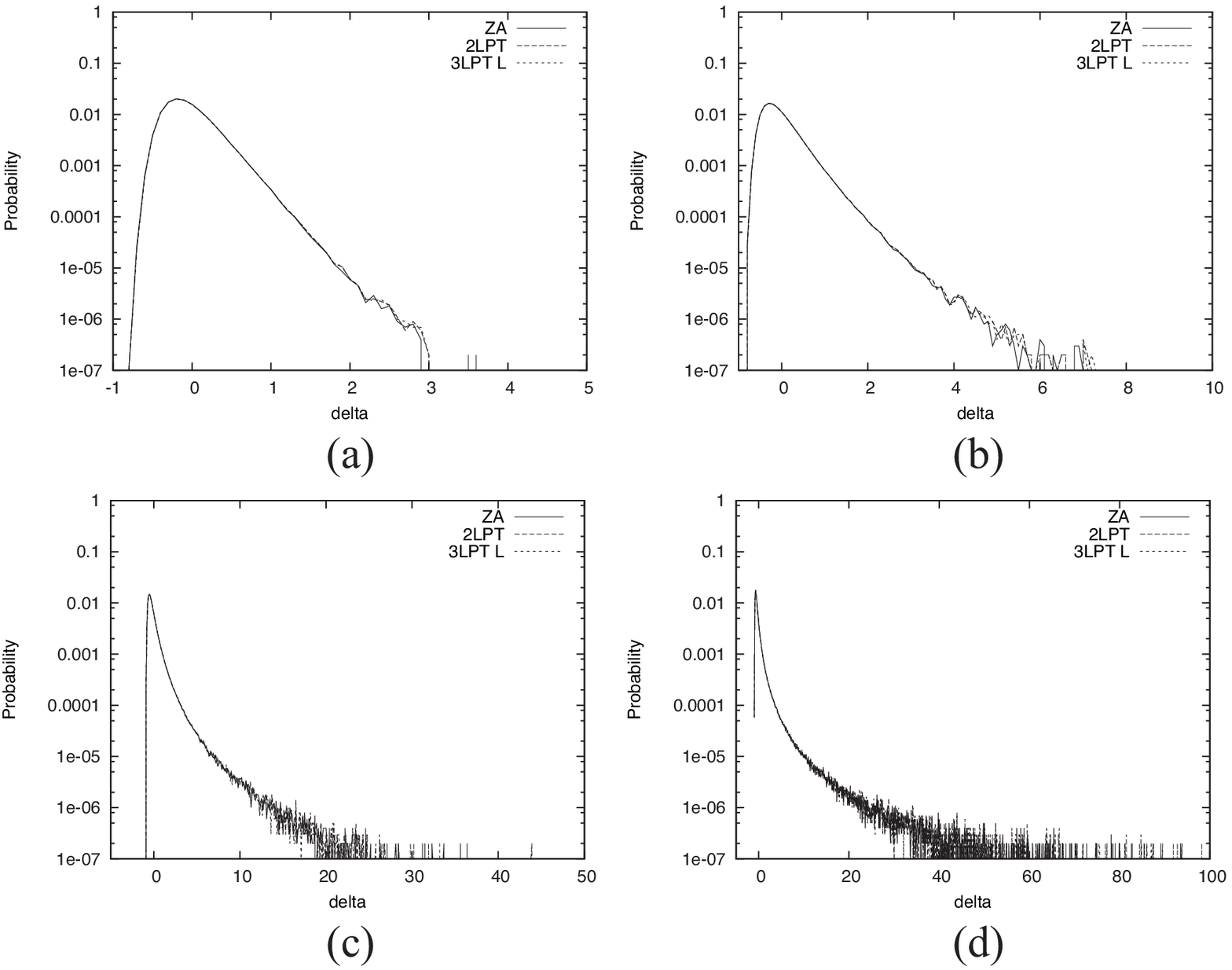}
}
\caption{Distribution function of the density fluctuation
from the $N$-body simulation ($R \simeq 2 h^{-1}$ Mpc) with different
initial conditions. 
(a) $z=5$, (b) $z=3$, (c) $z=1$, (d) $z=0$.
The distribution function approaches the log-normal form during the evolution.
}
\label{fig:delta-dist-2M}
\end{figure}

We change the smoothing scale to $R \simeq 2 h^{-1}$ [Mpc]. Even if
the smoothing scale is changed, the tendency of 
the distribution function for the density fluctuation is similar to the
case of $R \simeq 1 h^{-1}$ [Mpc]. The distribution function is
shown in Fig.~\ref{fig:delta-dist-2M}. 

\begin{figure}[tb]
\centerline{
\includegraphics[height=5cm]{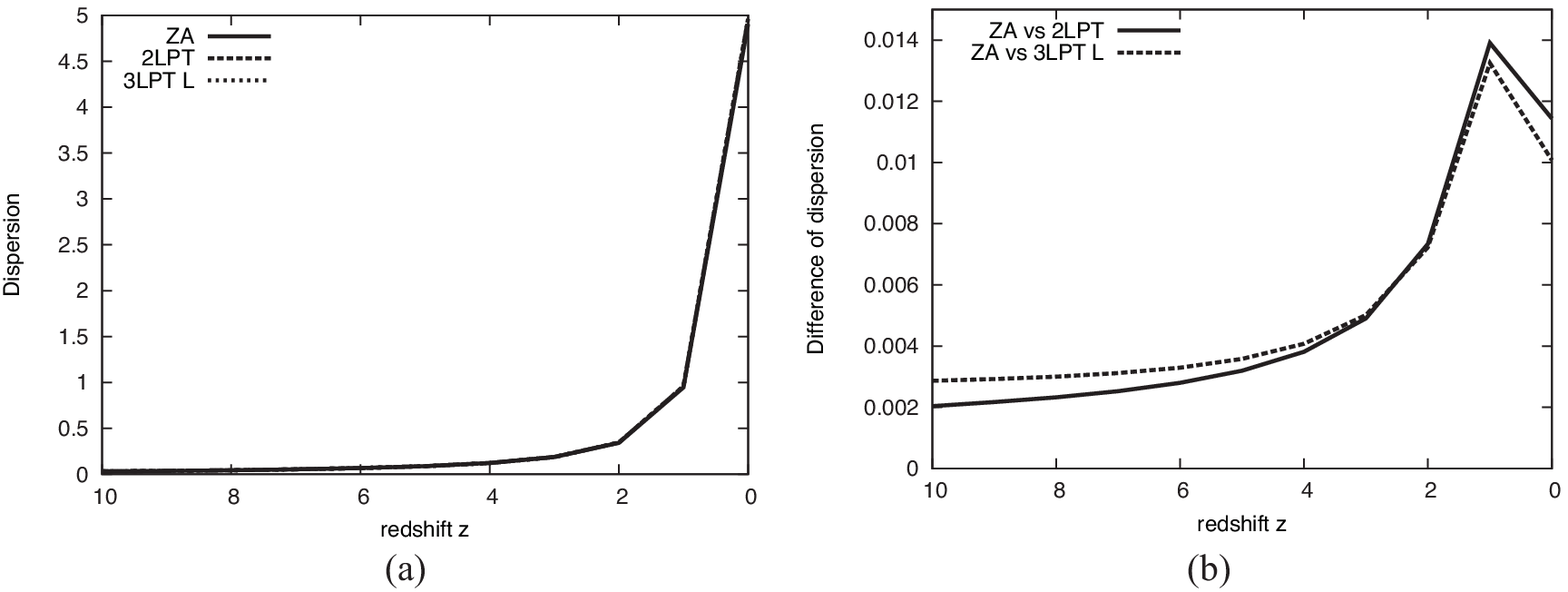}
}
\caption{Dispersion of the density fluctuation
from the $N$-body simulation ($R \simeq 2 h^{-1}$ Mpc) with different
initial conditions. (a) Comparison of the dispersion
between the initial conditions. (b) The relative difference in the dispersion
between ZA and other cases.
}
\label{fig:dispersion2M}
\end{figure}

\begin{figure}[tb]
\centerline{
\includegraphics[height=5cm]{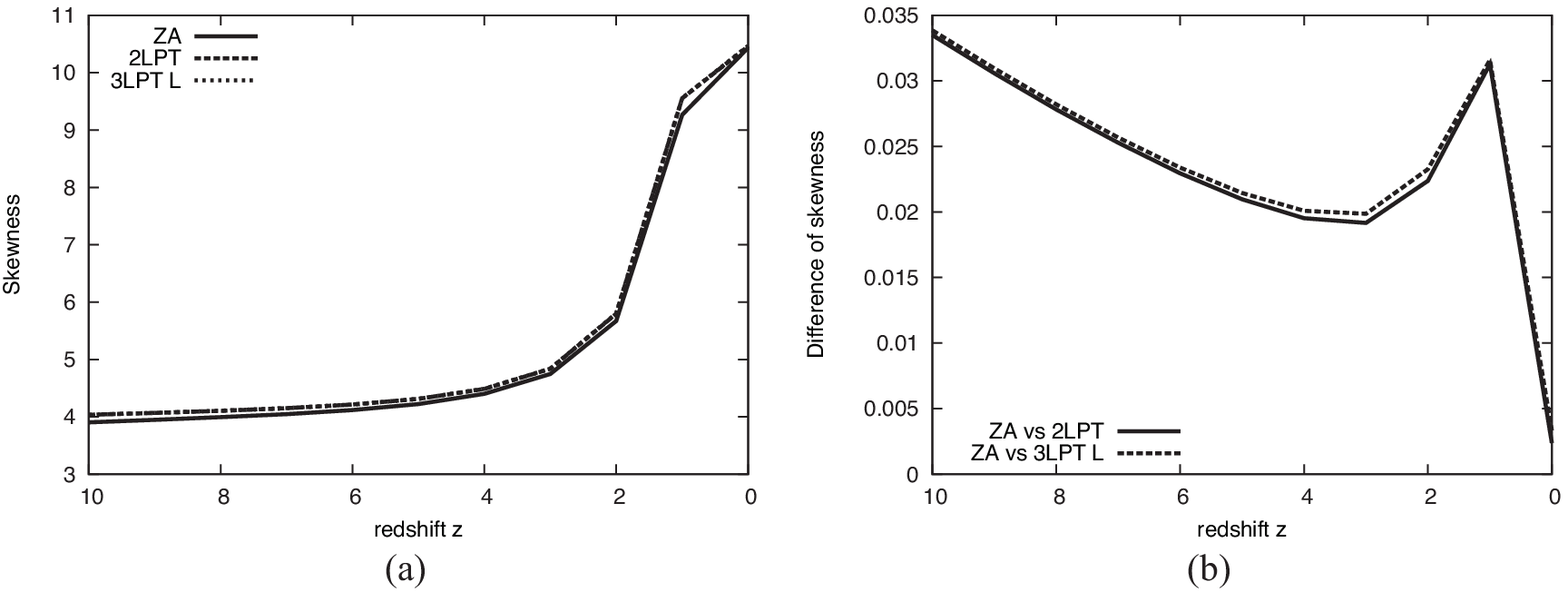}
}
\caption{Skewness of the density fluctuation
from the $N$-body simulation ($R \simeq 2 h^{-1}$ Mpc) with different
initial conditions. (a) Comparison of the skewness
between the initial conditions. (b) The relative difference in the skewness
between ZA and other cases.
}
\label{fig:skewness2M}
\end{figure}

\begin{figure}[tb]
\centerline{
\includegraphics[height=5cm]{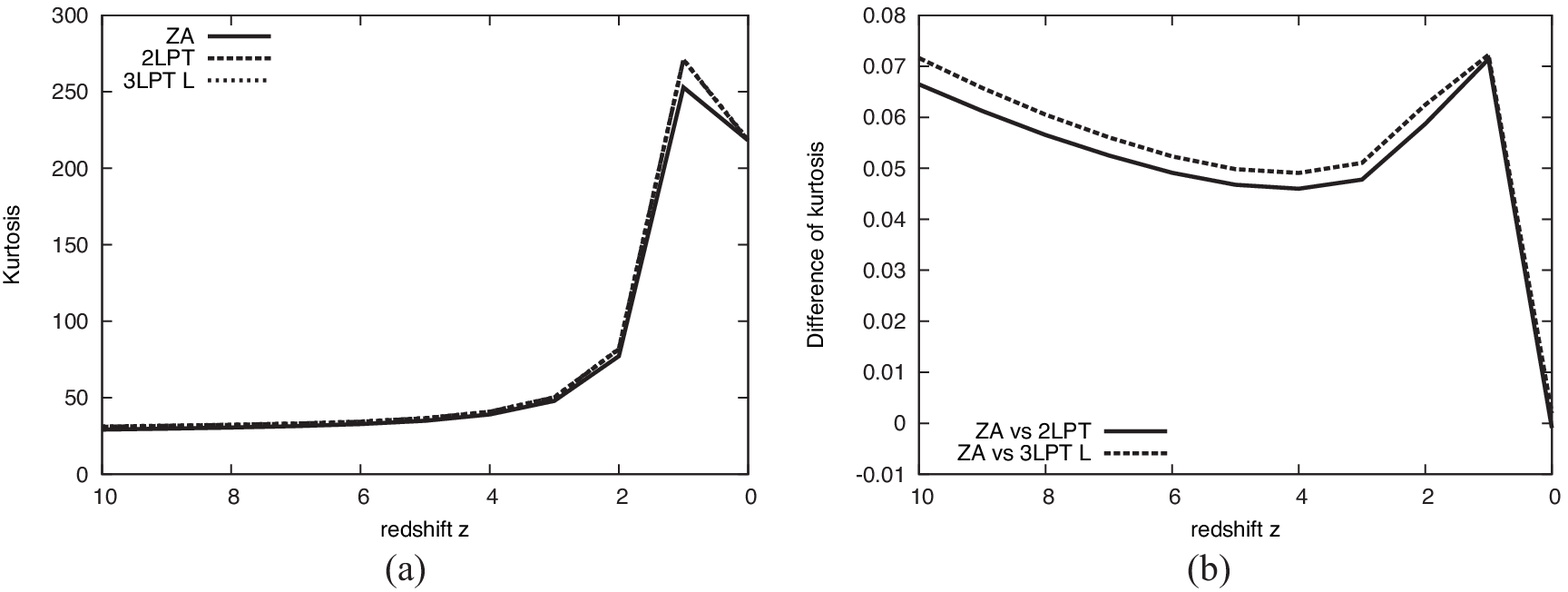}
}
\caption{Kurtosis of the density fluctuation
from the $N$-body simulation ($R \simeq 2 h^{-1}$ [Mpc]) with different
initial conditions. (a) Comparison of the kurtosis
between initial conditions. (b) The relative difference in the kurtosis
between ZA and other cases.
}
\label{fig:kurtosis2M}
\end{figure}

Time evolution of dispersion of the density fluctuation is shown 
in Fig.~\ref{fig:dispersion2M}. The non-Gaussianity of the distribution of
the density fluctuation is shown in Figs.~\ref{fig:skewness2M}
and ~\ref{fig:kurtosis2M}.
The difference of skewness and kurtosis between the case of
2LPT and 3LPT is about $0.1 \%$ and $0.3 \%$, respectively.

\begin{figure}[tb]
\centerline{
\includegraphics[height=10cm]{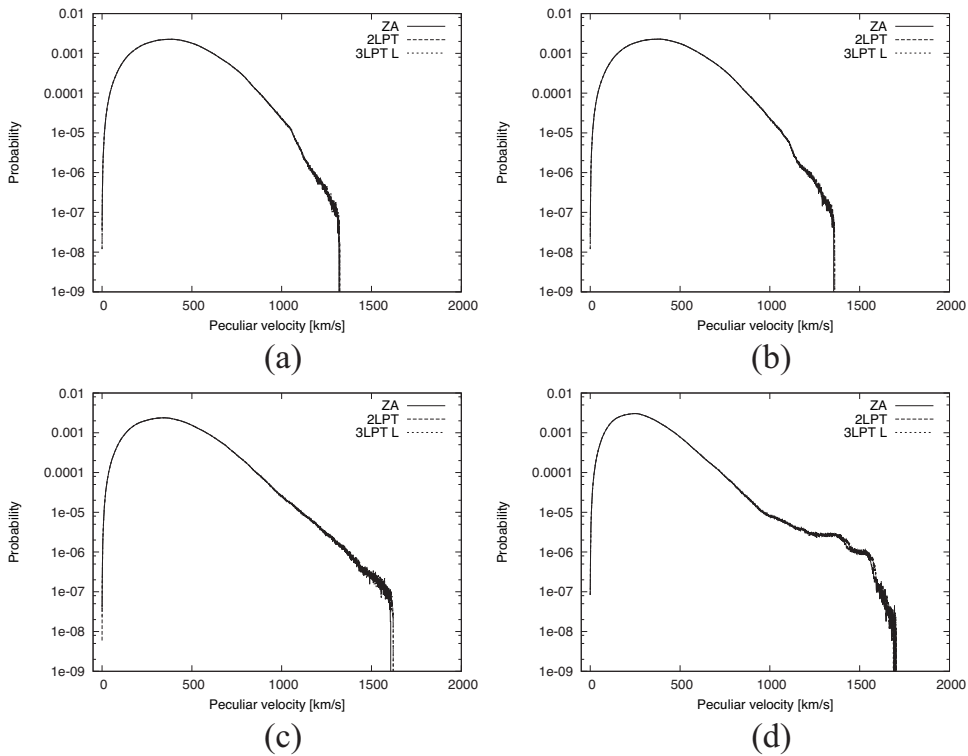}
}
\caption{The distribution function of the peculiar velocity
from the $N$-body simulation with different
initial conditions. 
(a) $z=5$, (b) $z=3$, (c) $z=1$, (d) $z=0$.
At high-z era, the effect of higher-order perturbation in the initial
conditions appears.
}
\label{fig:vel-dist}
\end{figure}

\begin{figure}[tb]
\centerline{
\includegraphics[height=11cm]{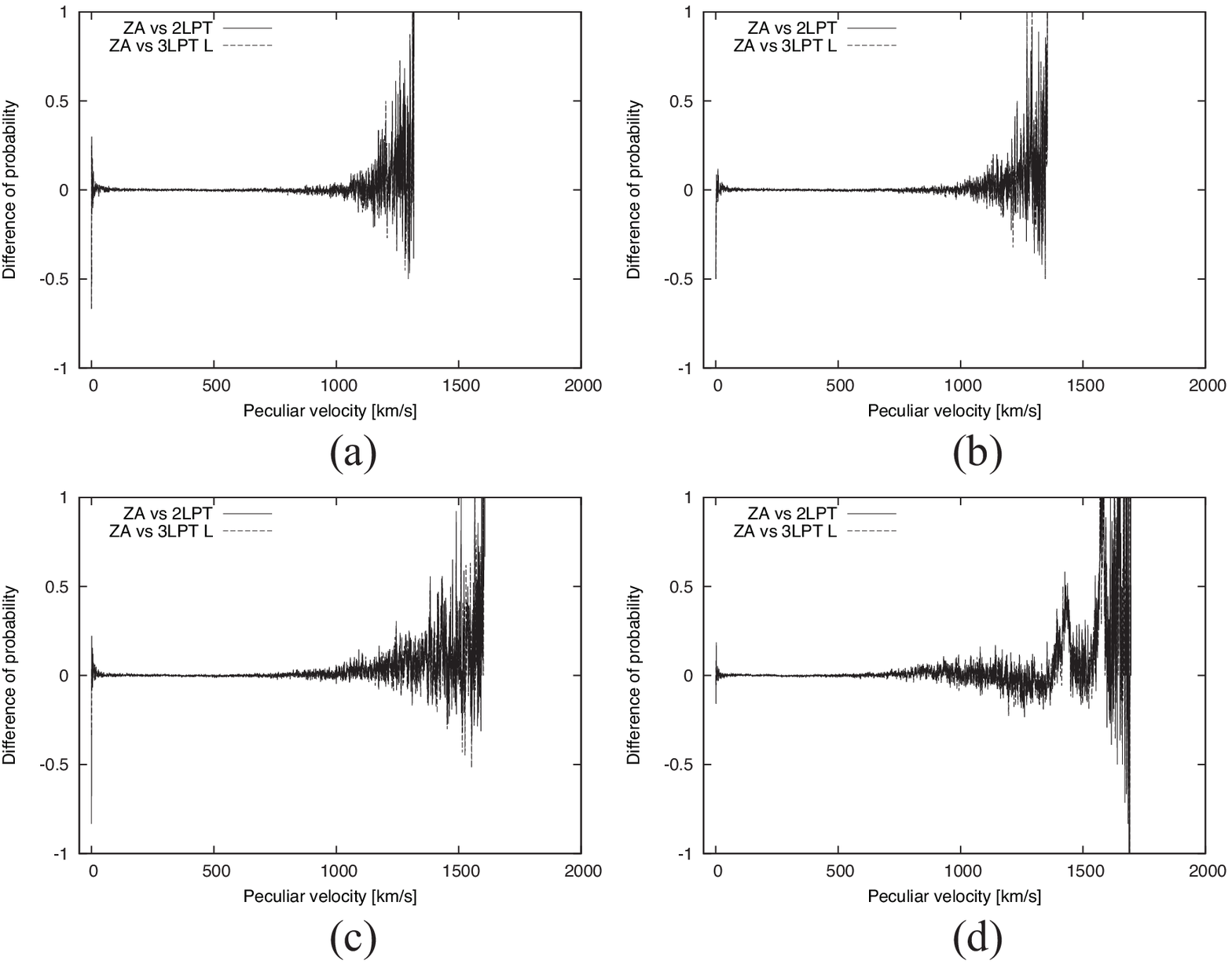}
}
\caption{The relative difference in the distribution function of the peculiar velocity
between the case of ZA and other cases.
(a) $z=5$, (b) $z=3$, (c) $z=1$, (d) $z=0$. }
\label{fig:vel-dist-diff}
\end{figure}

We noticed a distribution of peculiar velocity. Here, we compute the absolute 
value of the peculiar velocity for each particle. When clusters are formed,
the particles in clusters slow down. Therefore, the peculiar velocity
does not increase monotonically. The probability of the peculiar velocity
for each time is shown in Fig.~\ref{fig:vel-dist}. The effect of 
higher-order perturbation in the initial conditions appears in fast particles. 
For a more detailed analysis, we compare the probability of the peculiar
velocity between the case of ZA and other cases. The difference in
the probability of the peculiar velocity is shown in Fig.~\ref{fig:vel-dist-diff}.
The effect of higher-order perturbation appears in fast particles. When
we consider higher-order perturbation for the initial conditions of
the $N$-body simulation, the peculiar velocity increases.
At $z=5$, the number of fast particles (faster than $500$ [km/s]) in the case of 2LPT
is more than that in the case of ZA.
Similarly, the number of fast particles in the case of 3LPT is more than that in the case of 2LPT.
Although the particles form clusters at low-$z$ era ($z=3, 1, 0$),
the tendency continues afterwards. 
It was found that higher-order perturbations increased the number of
fast particles, but the number of fast particles was small
compared to the whole. Even if the density distribution is
considered in the redshift space instead of the real space,
the effect of higher-order perturbations on the deformation by
the peculiar velocity would be small.

\subsection{Dependence on initial time}
\begin{figure}[tb]
\centerline{
\includegraphics[height=5cm]{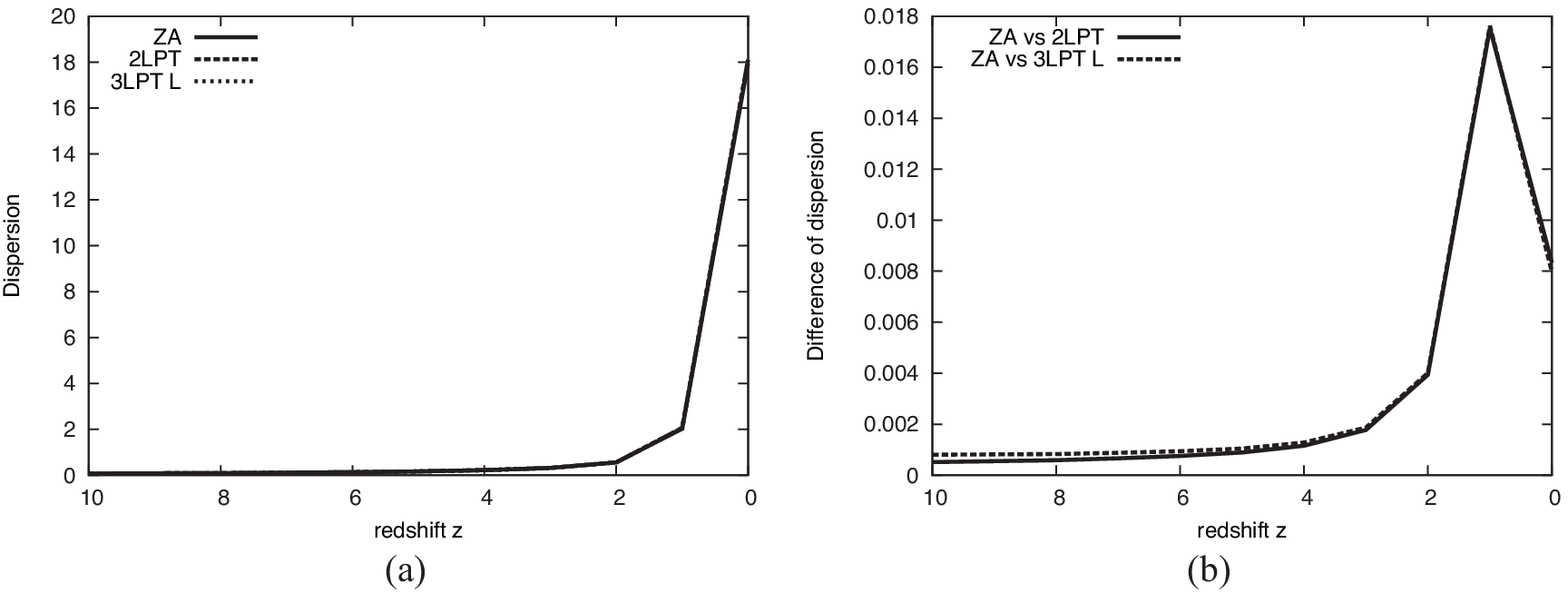}
}
\caption{Dispersion of the density fluctuation
from $N$-body simulation ($R \simeq 1 h^{-2}$ Mpc, $z_{\rm ini}=99$)
with different
initial conditions. (a) Comparison of the dispersion
between the initial conditions. (b) The relative difference of the dispersion
between ZA and other cases.}
\label{fig:dispersion-z99}
\end{figure}

The effect of higher-order perturbations increases during time
evolution. If the initial conditions for $N$-body simulation
set on early stage, will the effect of higher-order perturbation
weaken? To verify this conjecture,
we change the initial time for $N$-body simulation
to $z=99$. 
In this simulation, we generated 10 initial conditions for each case
(ZA, 2LPT, and 3LPT L).
The way of allocating pseudo-random numbers when creating the
initial conditions is the same as that for cases of $z=49$.
Fig.~\ref{fig:dispersion-z99} shows the evolution
of the density dispersion. Compared to the case of $z=49$,
the growth of dispersion is slightly slower for the case of $z=99$.

\begin{figure}[tb]
\centerline{
\includegraphics[height=5cm]{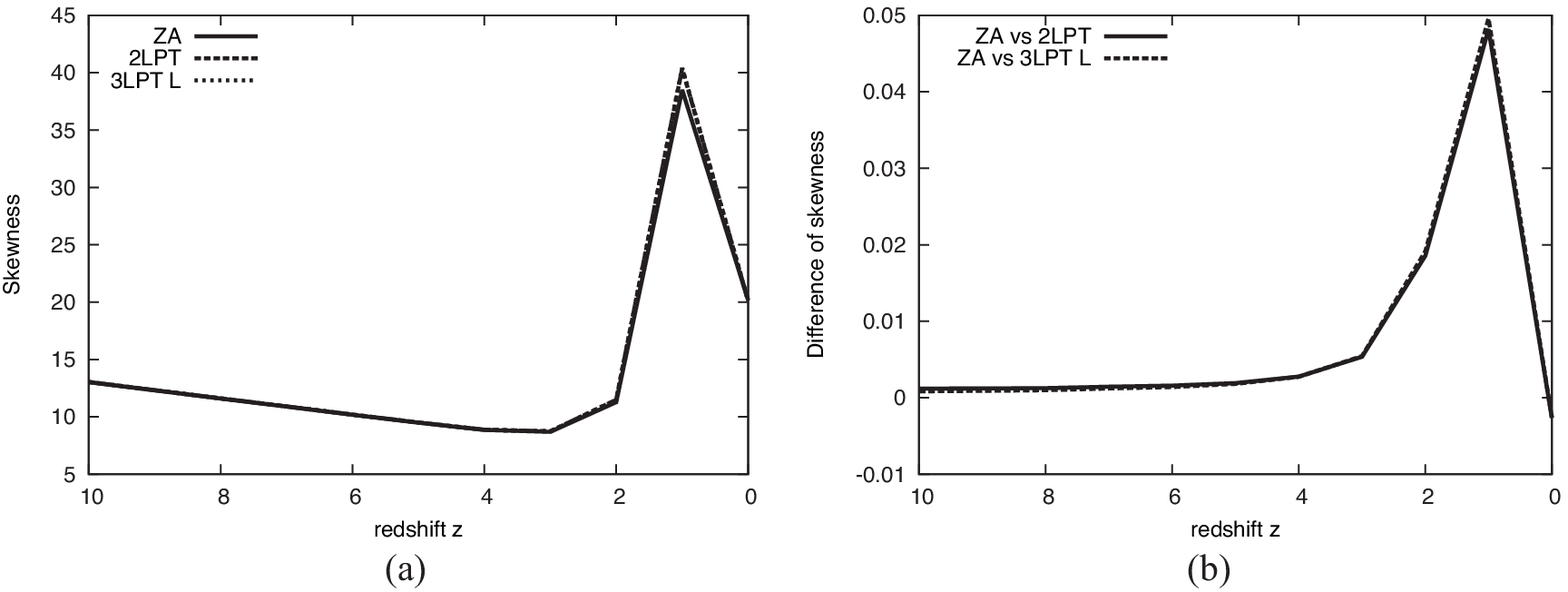}
}
\caption{Skewness of the density fluctuation
from the $N$-body simulation ($R \simeq 1 h^{-1}$ Mpc, $z_{\rm ini}=99$)
with different
initial conditions. (a) Comparison of the skewness
between the initial conditions. (b) The relative difference in the skewness
between ZA and other cases.
}
\label{fig:skewness-z99}
\end{figure}

\begin{figure}[tb]
\centerline{
\includegraphics[height=5cm]{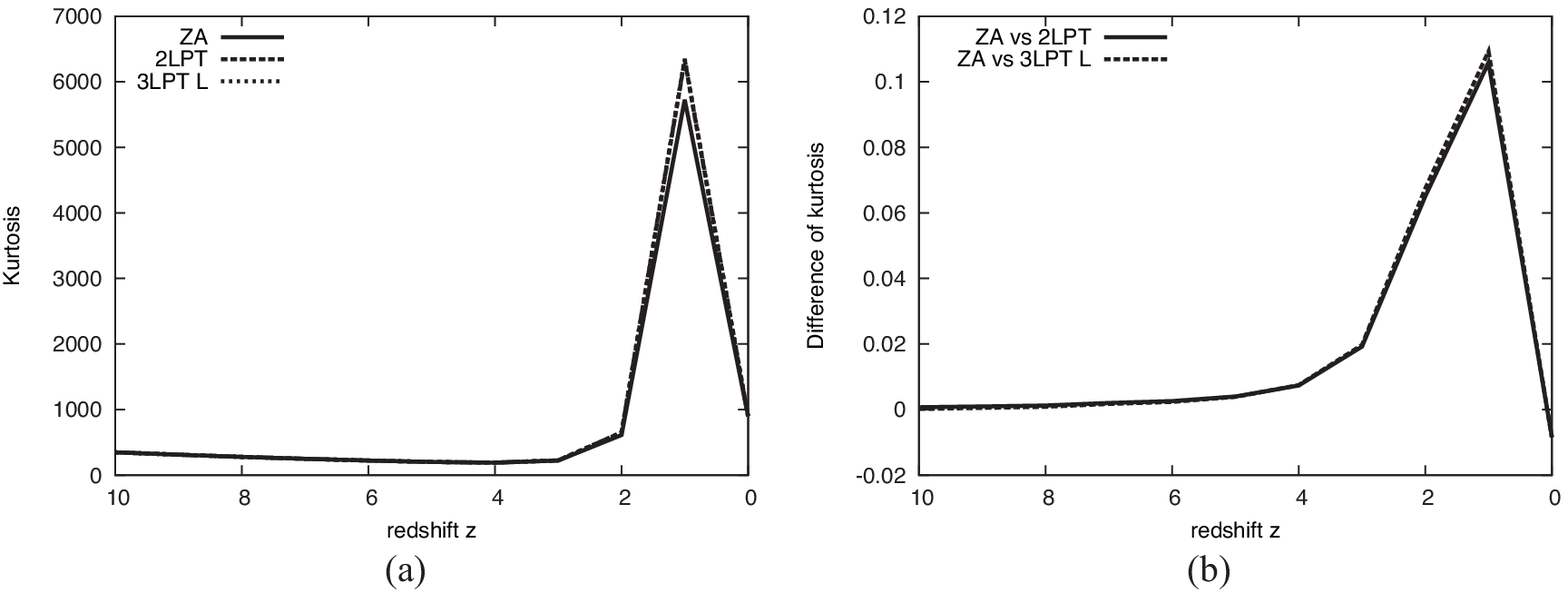}
}
\caption{Kurtosis of the density fluctuation
from the $N$-body simulation ($R \simeq 1 h^{-1}$ [Mpc], $z_{\rm ini}=99$)
 with different
initial conditions. (a) Comparison of the kurtosis
between the initial conditions. (b) The relative difference in the kurtosis
between ZA and other cases.
}
\label{fig:kurtosis-z99}
\end{figure}

We show the evolution of the non-Gaussianity in
Figs.~\ref{fig:skewness-z99} and \ref{fig:kurtosis-z99}. 
By comparison between the case of ZA and higher-order perturbations,
the difference in the skewness and the kurtosis is less than $5\%$ and
$12 \%$, respectively.
Compared to the case of $z_{\rm ini}=49$, the difference between the
models is smaller when $z_{\rm ini}=99$, but the non-Gaussian
difference between ZA and 2LPT remains several percents.
By comparison between the case of 2LPT and 3LPT,
the difference in the kurtosis is less than $0.5\%$.
In this analysis, the effect of higher-order perturbations
in the initial conditions disappears in the non-Gaussianity 
at high-z era ($z \simeq 10$). On the other hand,
the effect remains in the non-Gaussianity 
at low-z era ($z < 2$). 

\subsection{Dependence on number of particles}
\begin{figure}[tb]
\centerline{
\includegraphics[height=5cm]{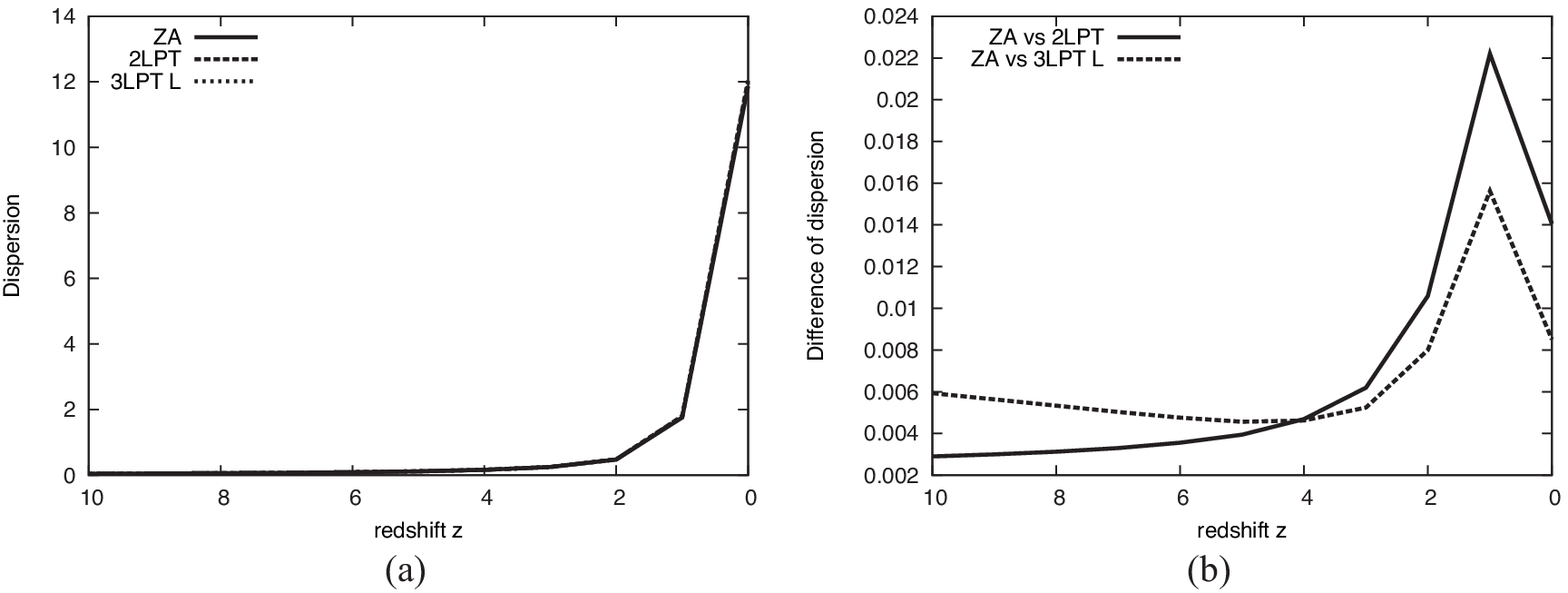}
}
\caption{Dispersion of the density fluctuation
from N-body simulation ($R \simeq 1 h^{-2}$ Mpc, $N=512^3$)
with different
initial conditions. (a) Comparison of the dispersion
between the initial conditions. (b) The relative difference of the dispersion
between ZA and other cases.}
\label{fig:dispersion-N512}
\end{figure}

For cosmological $N$-body simulations,  
the characteristics of the formed structure may be affected
by the difference in the number of particles.
In this paper, in order to avoid the effect of the number
of particles, we perform simulations with a large number
of particles. The number of particles is increased to $N=512^3$. 
Since this simulation takes a long time (about 60 hours),
only 7 samples were executed in this paper.
The other parameters are the same as those listed in Table
~\ref{tab:cosmo-param} and \ref{tab:Gadget-param}.
In this simulation, we generated 7 initial conditions for each case
(ZA, 2LPT, and 3LPT L).
Fig.~\ref{fig:dispersion-N512} shows the evolution
of the density dispersion. Compared to the case of $N=256^3$,
Compared to the case of $N=256^3$, the tendency of evolution is similar
in the case of $N=512^3$. The difference in the density dispersion
between models is similar for cases of $N=512^3$ and $N=256^3$.
By comparison between the case of ZA and higher-order perturbations,
the difference in the density dispersion is about $10\%$.
The difference of the density dispersion between the case of
2LPT and 3LPT is about $1 \%$.
The variation in the density dispersion appears because the number of
samples is small.

\begin{figure}[tb]
\centerline{
\includegraphics[height=5cm]{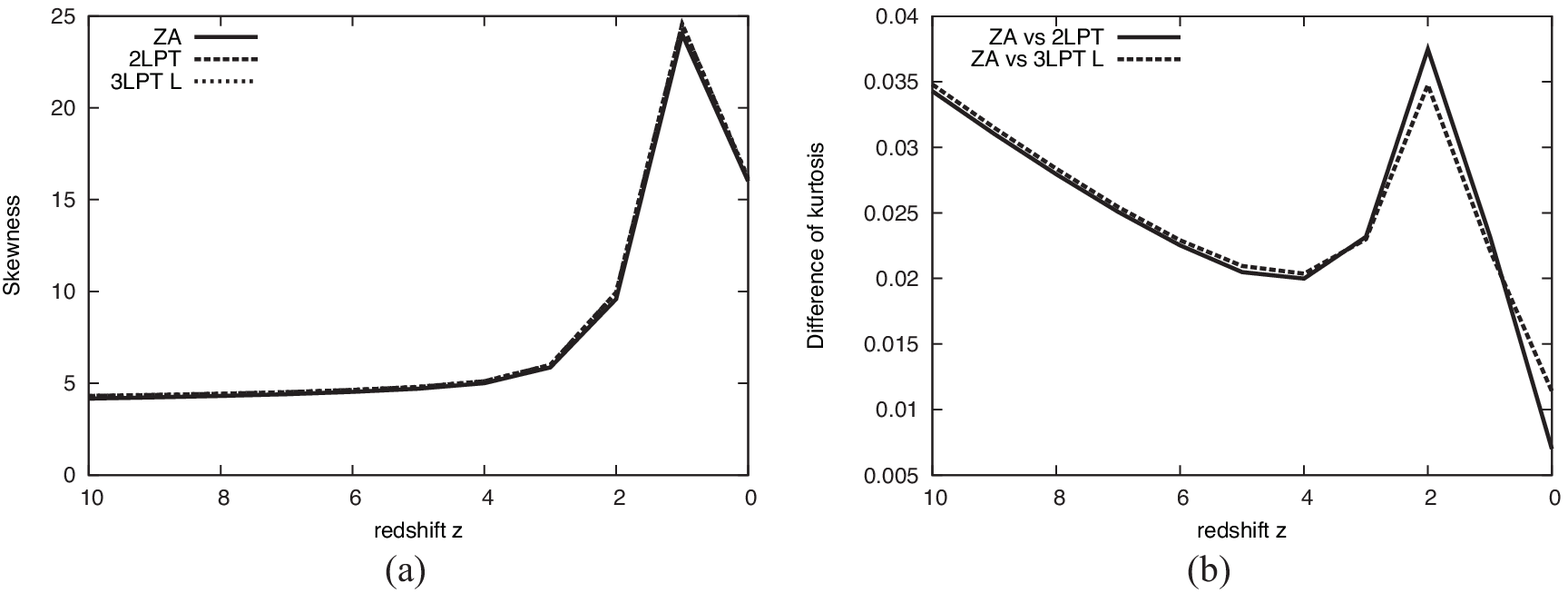}
}
\caption{Skewness of the density fluctuation
from the N-body simulation ($R \simeq 1 h^{-1}$ Mpc, $N=512^3$)
with different
initial conditions. (a) Comparison of the skewness
between the initial conditions. (b) The relative difference in the skewness
between ZA and other cases.
}
\label{fig:skewness-N512}
\end{figure}

\begin{figure}[tb]
\centerline{
\includegraphics[height=5cm]{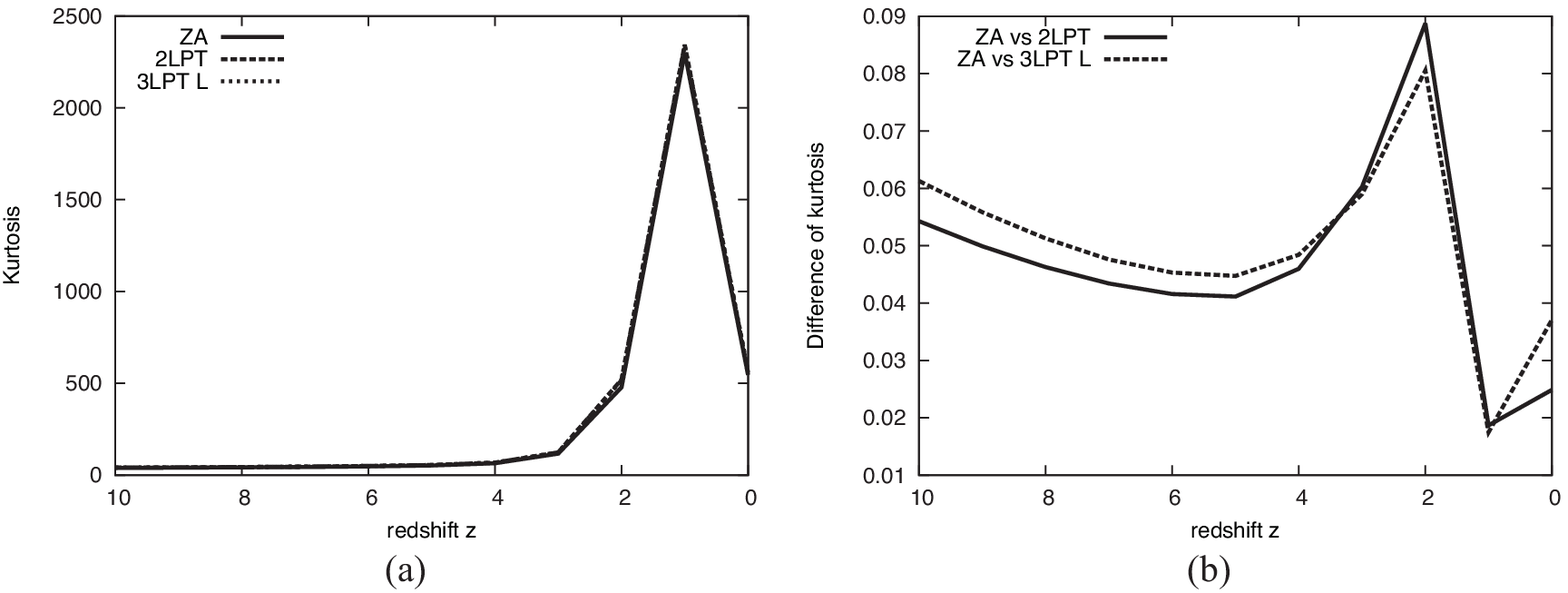}
}
\caption{Kurtosis of the density fluctuation
from the N-body simulation ($R \simeq 1 h^{-1}$ [Mpc], $N=512^3$)
 with different
initial conditions. (a) Comparison of the kurtosis
between the initial conditions. (b) The relative difference in the kurtosis
between ZA and other cases.
}
\label{fig:kurtosis-N512}
\end{figure}

We show the evolution of the non-Gaussianity in
Figs.~\ref{fig:skewness-N512} and \ref{fig:kurtosis-N512}. 
In the case of $N=512^3$, the same evolution tendency as in the
case of $N=256^3$ can be seen. Regarding the evolution of the
non-Gaussianity, the differences between the models are similar for
cases of $N=512^3$ and $N=256^3$.
By comparison between the case of ZA and higher-order perturbations,
the difference in the non-Gaussianity is about $10\%$.
When $z = 0$, the difference in the non-Gaussianity
between the cases of 2LPT and 3LPT becomes very small.
The difference of skewness and kurtosis between the case of
2LPT and 3LPT is about $0.4 \%$ and $1 \%$, respectively.
The variation in non-Gaussianity appears because the number of
samples is small.

From the above results, it was shown that the effect of
higher-order perturbations in the initial condition is almost
independent of the number of particles.

\subsection{Effect of the transverse mode in 3LPT} \label{sec:3LPT-T}
\begin{figure}[tb]
\centerline{
\includegraphics[height=10cm]{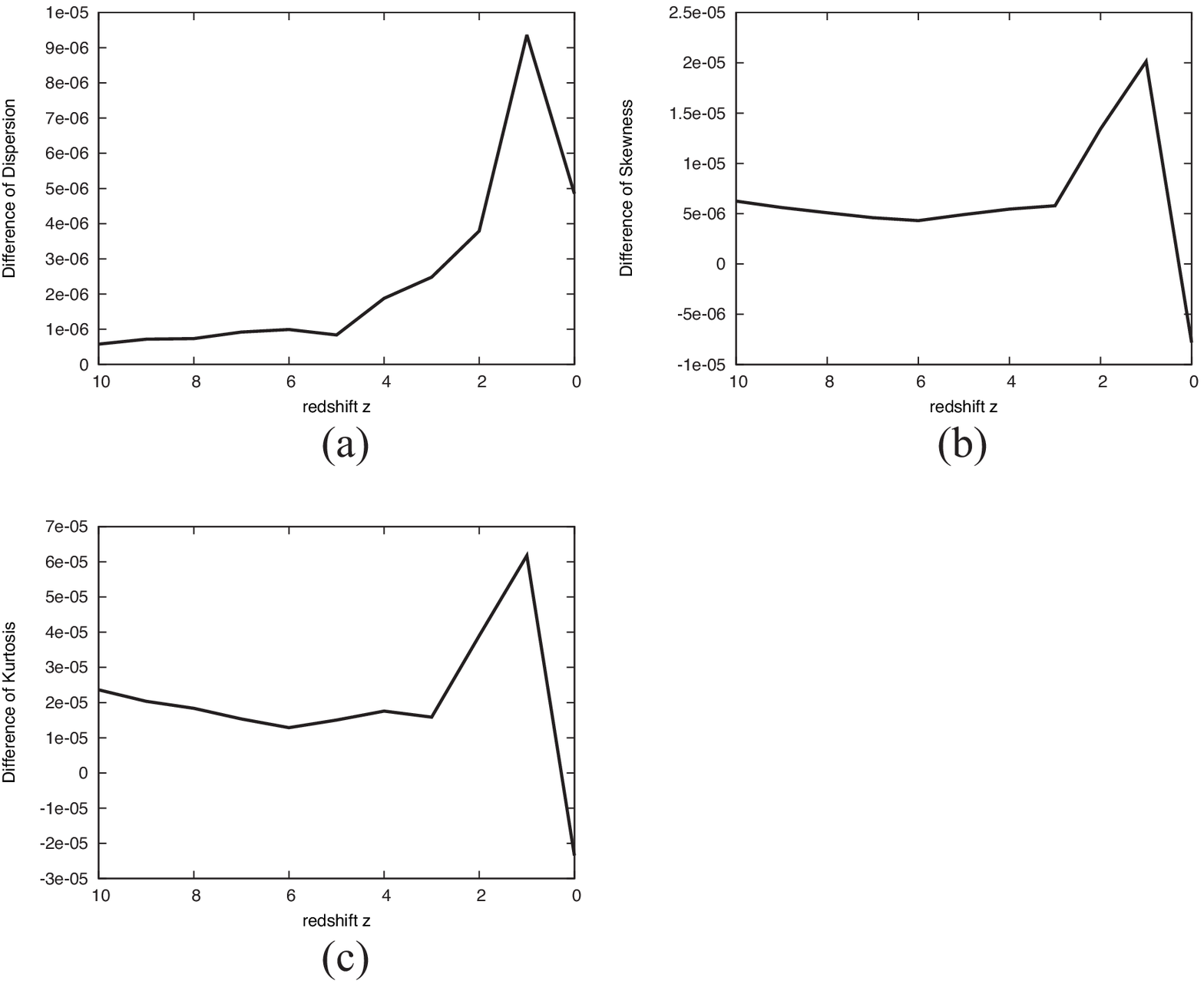}
}
\caption{The relative difference of non-Gaussianity with and without 
the transverse mode
($R \simeq 1 h^{-1}$ [Mpc]). This figure shows 
how much it shifts when the transverse mode is included.
(a) Dispersion, (b) skewness, (c) kurtosis.
}
\label{fig:trans-ng-diff}
\end{figure}

We examine how the presence or absence of the transverse mode in 3LPT affects 
nonlinear structure. We compare the non-Gaussianity of the density fluctuation
with and without the transverse mode in 3LPT
(``3LPT L'' vs ``3LPT L+T''). Figure~\ref{fig:trans-ng-diff}
shows how much the
non-Gaussianity of the density fluctuation that includes the transverse mode
deviates from the one that does not include the transverse mode. In any case, 
the difference between 2LPT and 3LPT is much smaller and negligible.

The reason can be explained using Eq.~(\ref{eqn:delta-Jacobian}). 
Eq.~(\ref{eqn:delta-Jacobian}) can be rewritten as follows.
\begin{eqnarray}
\delta &=& - \left ( s_{i,i} +
 \frac{1}{2} \left ( s_{i,i} s_{j,j} - s_{i,j} s_{i,j} \right )
  + \det \left ( s_{i,j} \right ) \right ) \nonumber \\
  && \cdot  \left ( 1 + s_{i,i} +
 \frac{1}{2} \left ( s_{i,i} s_{j,j} - s_{i,j} s_{i,j} \right )
  + \det \left ( s_{i,j} \right ) \right )^{-1} \,.
\end{eqnarray}
This equation is expanded to the order of perturbation.
\begin{eqnarray}
\delta &=& - \psi_{,ii}^{(1)} \nonumber \\
 && + \left [ - \psi_{,ii}^{(2)} + 
 \frac{1}{2} \left ( \psi_{,ii}^{(1)} \psi_{,jj}^{(1)} 
 + \psi_{,ij}^{(1)} \psi_{,ij}^{(1)} \right ) \right ] \nonumber \\
 && + \left [ - \psi_{,ii}^{(3)} + 
  \left ( \psi_{,ii}^{(1)} \psi_{,jj}^{(2)} 
 + \psi_{,ij}^{(1)} \psi_{,ij}^{(2)} \right ) 
 - \psi_{,ii}^{(1)} \psi_{,jk}^{(1)} \psi_{,jk}^{(1)} 
 - \det \left ( \psi_{,ij}^{(1)} \right ) \right ] \nonumber \\
 && + \left [ \psi_{,ij}^{(1)} \left ( \psi_{,ij}^{(3)} + \zeta_{i,j}^{(3)} \right )
  + F_4 \left ( \psi_{,ij}^{(1)},  \psi_{,ij}^{(2)}, \psi_{,ij}^{(3)} \right )
  \right ] \nonumber 
 + O (\varepsilon^5) \,,
\end{eqnarray}
where $F_4$ means fourth-order perturbative quantity
which composed of first-, second-, and third-order perturbation
in the longitudinal mode. $O (\varepsilon^5)$ means
fifth- or more higher-order perturbative quantities. 
The transverse mode appears in
fourth-order term. Therefore, at least the transverse mode has only 
a fourth-order or higher-order effect on density fluctuations.

\begin{figure}[tb]
\centerline{
\includegraphics[height=5cm]{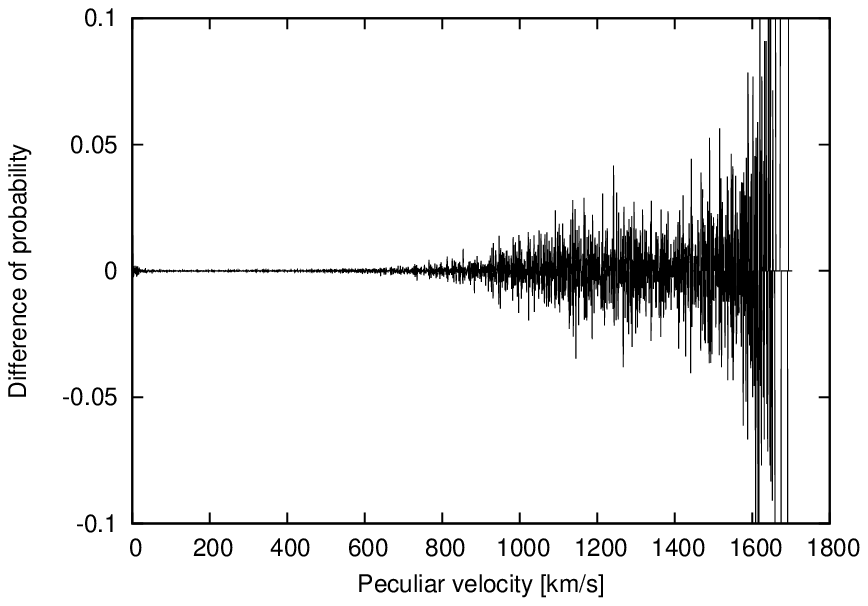}
}
\vspace{15mm}
\caption{The relative difference in the distribution function of the peculiar velocity
between the presence and absence of the transverse mode ($z=0$).
The probability distribution of fast velocities varies greatly
between them.
}
\label{fig:3LPT-vdist-diff}
\end{figure}

The effect of the transverse mode clearly appears on the velocity distribution. 
Figure~\ref{fig:3LPT-vdist-diff} shows how the difference
between the presence and absence of the transverse mode
in the initial conditions appears in the velocity distribution. 
The probability distribution of fast velocities varies greatly
between them.

\subsection{Comparison with past research}
\begin{figure}[tb]
\centerline{
\includegraphics[height=10cm]{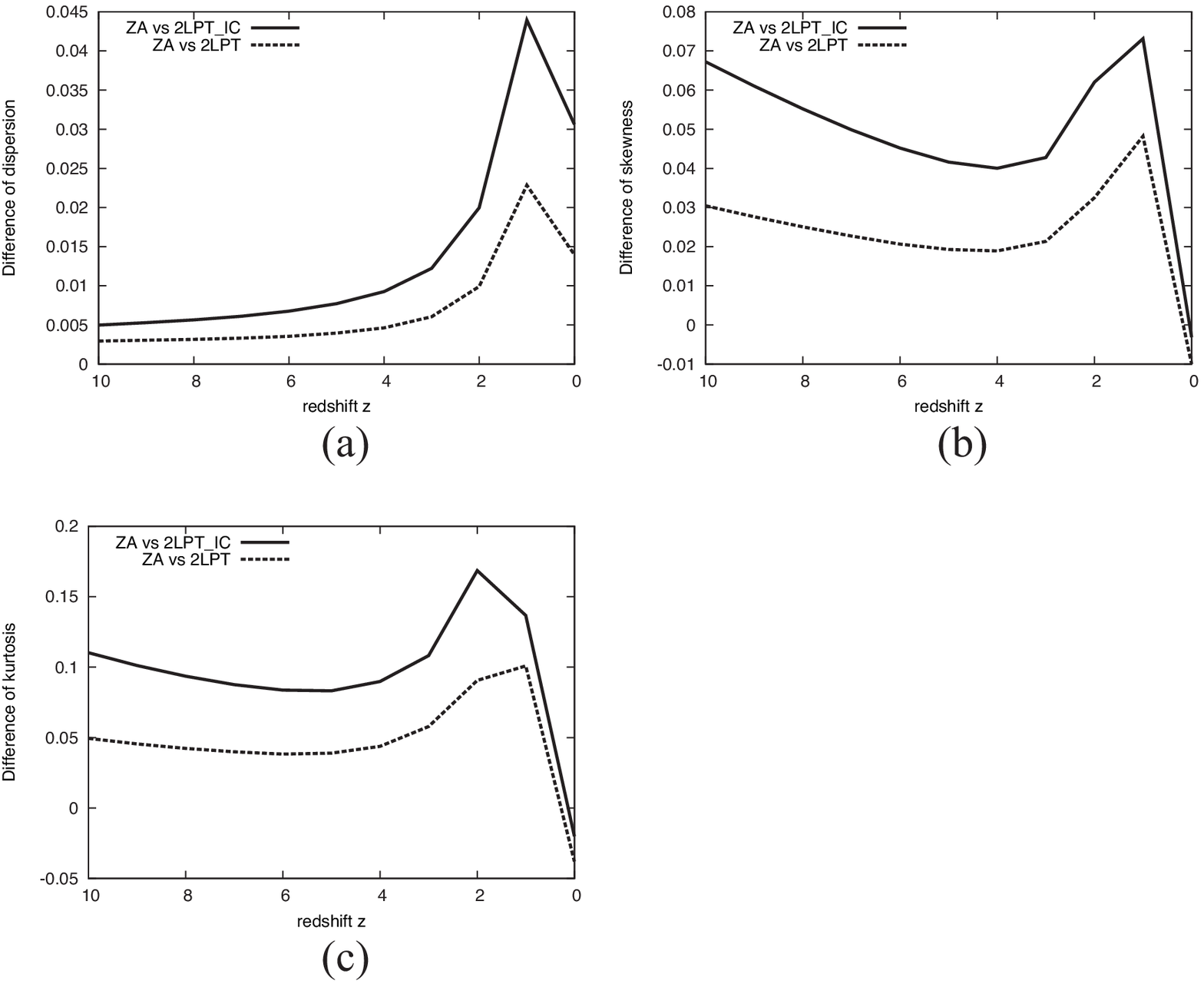}
}
\caption{The relative difference of the density distribution between
2LPT\_IC code and our code ($R \simeq 1 h^{-1}$ Mpc). 
This figure shows 
the difference between the case of ZA and each code.
(a) Dispersion, (b) skewness, (c) kurtosis.
}
\label{fig:2LPT-diff-ng}
\end{figure}

The validity of this study is compared with past studies. Therefore,
we compare our results with those using 2LPT\_IC code. In this paper,
we examine the non-Gaussianity of the density distribution between
2LPT\_IC code and our code. Figure~\ref{fig:2LPT-diff-ng} shows how much it deviates
from ZA. For each of these quantities, the case of 2LPT\_IC code
shows a slightly larger deviation than the case of our code. This result
shows a larger deviation than when the order of the perturbation was
raised to the third-order in our code.

The difference between the two codes is thought to depend on the
method used to solve Poisson equation. In 2LPT\_IC code, the derivative
of the perturbation is calculated in Fourier space to solve Poisson equation.
On the other hand, in our code, the derivative of the perturbation is
calculated in real space to solve Poisson equation. It is thought that
the difference between the two methods caused the difference. The
perturbation growth rate is not corrected by the density parameter
in our code, but it seems that this effect is small at $z_{\rm ini}=49$. On the
contrary, the growth of higher-order perturbations in our code is
larger than these in 2LPT\_IC code.


\section{Summary}\label{sec:Summary}
We analysed the effect of higher-order perturbation for the initial
conditions of cosmological $N$-body simulations. Based on our previous studies, 
we developed an initial condition converter for Gadget-2 code.

For density fluctuation, the effect of higher-order perturbation appeared
in strongly nonlinear region. Although the primordial density fluctuation
was generated by the Gaussian distribution, because of nonlinear evolution,
the non-Gaussianity of the distribution of the density fluctuation appeared
at low-$z$ era. Further, we compared the statistical quantities for
the non-Gaussianity. Although it varied significantly depending on the samples,
the effect of 3LPT on the initial condition was evident.
Even if the time of the initial condition is set early,
although the effect of 3LPT would disappear,
the effect of 2LPT affects the
evolution of the density distribution later.
It was also clarified that the effect of 
higher-order perturbations hardly depends on
the number of particles in the simulation.

With regard to the peculiar velocity, the higher-order perturbation affected fast
particles, and during clustering, the effect gradually disappeared.
Considering the density distribution in redshift space,
the effect of higher-order perturbation would appear in the shape
of structures such as finger-of-god~\cite{Jackson1972, Scoccimarro2004}.

For the 2LPT initial conditions, the results obtained using our code were slightly
different from that obtained by the 2LPT\_IC code. In our code, the spatial differential
was calculated by the difference in the Lagrangian space. On the other hand,
in the 2LPT\_IC code, the spatial differential was calculated in the Lagrangian 
Fourier space. Further, in our code,
the time evolution from recombination era to
the initial time ($z_{\rm ini}=49$) was given by the growing factor in the E-dS Universe mode.
In the 2LPT\_IC code, the time evolution was given by an approximated formula
for the $\Lambda$CDM model. The difference in the initial set up 
spread in the nonlinear stage.

This study shows that it is appropriate to apply initial conditions including 
2LPT when the accuracy of 3LPT is unnecessary.
Although the results obtained using both the codes varied slightly,
we demonstrated that the effect of 3LPT in the initial condition appeared in the nonlinear
stage. Therefore, for more precise prediction for a large-scale
structure (with $0.1 \%$ accuracy), the effect of 3LPT on the initial
condition should be considered
for cosmological $N$-body simulations. This effect would appear
in other statistical quantities as well. 

\section*{Acknowledgments}
We thank Shuntaro Mizuno and Toshihiro Nishimichi for useful comments.

\section{References}


\end{document}